\documentclass[aps,12pt,nofootinbib]{revtex4}
\usepackage{graphics}
\usepackage{graphicx}
\usepackage[dvips]{color}
\usepackage{amsmath}
\usepackage{amssymb}
\usepackage{aas_macros}

\def\beq{\begin{equation}}
\def\eeq{\end{equation}}

\def\bea{\arraycolsep .1em \begin{eqnarray}}
\def\eea{\end{eqnarray}}

\newcommand{\step}{\vspace{.5em}}

\def\s0#1#2{\mbox{\small{$ \frac{#1}{#2} $}}}
\def\0#1#2{\frac{#1}{#2}}

\def\grgl{\:\hbox to -0.2pt{\lower2.5pt\hbox{$\sim$}\hss}{\raise3pt\hbox{$>$}}\:}
\def\klgl{\:\hbox to -0.2pt{\lower2.5pt\hbox{$\sim$}\hss}{\raise3pt\hbox{$<$}}\:}

\def \gta {\mathrel{\vcenter
     {\hbox{$>$}\nointerlineskip\hbox{$\sim$}}}}

\newcommand \be {\begin{equation}}
\newcommand \ee {\end{equation}}
\newcommand \bed {\begin{displaymath}}
\newcommand \eed {\end{displaymath}}

\newcommand{\bit}{\begin{itemize}}
\newcommand{\eit}{\end{itemize}}


\begin{document}

\title{Isotropic Lifshitz critical behavior from the functional renormalization group}

\author{A.~Bonanno${}^{a,b}$ and D. Zappal\`a${}^{b}$}

\affiliation{
\mbox{${}^a$  INAF, Osservatorio Astrofisico di Catania, via S.Sofia 78, I-95123 Catania, Italy  }\\
\mbox{${}^b$ INFN, Sezione di Catania,  via S. Sofia 64, I-95123,
Catania, Italy.}
}%

\begin{abstract}
${}$\\[-1ex]

\centerline{\bf Abstract} 

The Lifshitz critical behavior  for a single component field theory
 is studied for the specific isotropic case in the framework of the 
 Functional Renormalization Group.  Lifshitz fixed point solutions of the flow equation,
 derived by using  a Proper Time regulator, are searched at  lowest and higher order in the 
 derivative expansion. Solutions are found when the number of spatial dimensions $d$ is contained 
 within the interval $5.5 < d < 8$.

\end{abstract}

\pagestyle{plain} \setcounter{page}{1}

\maketitle

\section{introduction}

Lifshitz critical points  represent a particular class of  tricritical points  on the phase diagram,
characterized  by  the coexistence of a disordered phase with vanishing order parameter,
a homogeneous ordered phase with finite constant order parameter, and a modulated phase
where the order parameter shows a periodic  structure with  finite wave vector,  
and a typical realization is observed in  a ferromagnet with  three phases:  paramagnetic, ferromagnetic 
and  helicoidal or sinusoidal.  

This idea  was first introduced  in \cite{horn}, by a generalization
of the usual Landau-Ginzburg $\phi^4$ model where the space coordinates of a $d$ dimensional
space are separated in parallel and orthogonal components, respectively spanning an $m$
and a $d-m$ dimensional space,  and the terms involving derivatives  with respect to one or 
to  the other set of coordinates in the action are treated differently.
Then, the usual kinetic term with the square gradient  of the field  in one set of coordinates can be 
kept finite  while  the square gradient related to the other set of coordinates  is suppressed, 
so that the dynamics of the terms with four powers of the gradient of the field becomes essential.

The anisotropy introduced between the two sets of coordinates  generates  
a multicritical point, namely the Lifshitz point, whose universal behavior is determined by 
the three parameters $(d,m,N)$ ($N$  indicates the number of components of the field 
$\phi$), and  which  shows a peculiar  critical behavior that  requires two different anomalous 
dimensions and two critical exponents, related to two different  correlation lengths, 
in order to  describe the  two-point correlation function. The interest in the Lifshitz points
is due to the large variety of systems that present this kind of critical behavior,
such as magnetic systems like the MnP compound or the so-called ANNNI model,
but also polymer mixtures, liquid crystals,  high-$T_c$ superconductors. For
reviews on this subject see \cite{hornrev,selke1988,diehl}.

More recently  the Lifshitz critical behavior  has found application in a theory of gravity formulated on the 
basis of a conjectured  anisotropy  between  time and  space  coordinates that  reduces
the ultraviolet pathologies of the theory \cite{horava,filippo,zz}.   
In addition, a duality between an $O(N)$ model at an isotropic Lifshitz point and a
higher spin gravity theory has been proposed in  \cite{bekaert1,bekaert2}, while  models involving
effects of Lorentz invariance violation due to a Lifshitz anisotropy are discussed
in \cite{alexandre,kikuchi}.

The critical properties of the Lifshitz point  were originally studied in the framework of the $\epsilon$-expansion \cite{horn},
and  the hard task of evaluating the free propagator 
for generic $m$ and $d$, made the calculation of  the $O(\epsilon^2)$ corrections a very difficult problem which was eventually
solved in \cite{carneiro,diehl1,diehl2}. Analogous difficulties appeared in the computation of the critical properties at 
large $N$ with the relative $O(1/N)$ corrections \cite{diehl4}.

Among the possible configurations of $d$ and $m$ for the Lifshitz critical point  there is  one, namely  
the case $m=d$, in which the isotropy is recovered again.  Then, all the space coordinates 
have the same critical behavior,  which however is different from the standard case where 
the kinetic term for all the space coordinates is quadratic in the gradient, $O(\partial^2)$,  while 
for $m=d$ it is quartic, $O(\partial^4)$.   The interest in the isotropic $m=d$ case
was primarily motivated  for its application to  the mixtures of  a  homopolymer blend and diblock copolymer
for which a Lifshitz point is predicted by mean field theory, and the measurement of 
critical exponents was performed in \cite{schwahn}.  The $\epsilon$ expansion in this case is realized
along the diagonal $m=d$ with $\epsilon=8-d$ \cite{horn,diehl3}.
More recently, a numerical approach by Monte Carlo simulations indicates that the isotropic Lifshitz points could be 
destroyed upon inclusion of fluctuations \cite{schmid}.
Therefore, it is certainly of interest  to  analyze the problem by means of  a different  non-perturbative approach, 
suitable to study systems at criticality, namely the Functional Renormalization Group (FRG).

The FRG approach, \cite{polch,wett1,morris1}, consists of a functional differential flow equation for the running effective action
which provides a description of the physics at an energy scale $k$ that, in turn, runs from a large ultraviolet scale, 
where the bare action is defined, down to small scales and eventually to $k=0$, where the running action 
becomes equal to  the standard effective action. The flow equation is the result of the progressive integration
of the fluctuations with momentum contained in an infinitesimal interval centered around $k$, 
so that,
when $k=0$, all fluctuations have been  integrated out.  In practice, the integration 
of the fluctuations  is performed by introducing a particular cut-off function that selects the desired interval of modes
and, clearly, the flow equation carries  an explicit dependence  on the specific cut-off  employed. 
Many reviews are available on the various  formulations of the FRG flow equation and on its
numerous applications \cite{bagnuls,wett2,poloni,lit1,pawl}.

The idea of applying the FRG equation to the study of the Lifshitz critical behavior is not new,
as it is implemented in \cite{bervillier}, where the Lifshitz fixed  point  and the main critical exponents are 
evaluated in the uniaxial case $m=1$, and in \cite{essafi},  where the specific  case 
of $N=3$ and $m=1$ is analyzed. Both papers show that the FRG approach is suitable 
to investigate these critical properties, avoiding some  technical difficulties encountered 
within other non-perturbative approaches.
However the FRG has not yet been used in a numerical  study  of the isotropic case
$m=d$. 

Therefore, in this paper we focus on the isotropic $m=d$ problem and make use of the  
Proper Time  (PTRG) version of the FRG equation.  
This is a flow equation  originally derived from a proper-time regularization of the one-loop effective action, \cite{liao,bohr,boza2001}
that can equally be obtained from generalized Callan-Symanzik flows \cite{litpaw2}, or, more generally can be derived 
in the framework of the background field flows \cite{litpaw3,litpaw4} .  
This equation, that has previously been used for studies 
of phase transitions \cite{litpaw2,bohr,boza2001,maza,litpaw5,bola,litimzappala},
spontaneous symmetry breaking and tunneling phenomena \cite{zap1,coza,zap2} , gravity \cite{Bonanno:1996ir,boro},  
has the advantage of being accurate and rapidly converging in the determination
of the critical properties of the theory and therefore suitable to approach the problem considered here.
In addition, as the  PTRG has been used in \cite{litimzappala} to study the Ising universality class to fourth order in the 
derivative expansion, thus  including the  $O(\partial^4)$ terms , we can take  advantage  of using the formalism 
already developed in \cite{litimzappala} to  study the Lifshitz critical point  for the one component field theory, $N=1$,
in the isotropic limit $m=d$.

In Sect. II, we recall the essential properties of the Lifshitz critical behavior; in Sect. III the PTRG and the structure of the 
corresponding flow equations are outlined, while  the numerical results are discussed in Sect. IV. Conclusions 
are summarized in Sect. V.

\section{Lifshitz critical behavior}

The general form  of a $d$-dimensional  action, suitable to investigate on the occurrence of a tricritical Lifshitz point is
\be
\Gamma [\phi]=\int {\rm d}^{d-m}x_\bot \; {\rm d}^{m}x_\parallel \left \{ W_\parallel\,(\partial_\parallel^2 \phi)^2  
+  W_\bot\, (\partial_\bot^2 \phi)^2  + \frac{ Z_\parallel } {2}  (\partial_\parallel \phi)^2 
+ \frac{  Z_\bot }{2} (\partial_\bot \phi)^2 +V(\phi) \right \}
\label{startaction}
\ee
where in general $\phi(x)$ is an $N$-component vector field, although
here we shall focus on the single component field theory with $N=1$.
The potential $V$ is a generic function of the field, while  
the coordinates $x$ are  decomposed in  parallel, $x_\parallel$, and  orthogonal, $x_\bot$, components, that respectively 
belong to an $m$-dimensional  and a $(d-m)$-dimensional subspace which possess two different scaling behaviors.
In fact, at mean field level, one observes that, by keeping $Z_\parallel > 0$, a vanishing and a non-vanishing minimum of 
the potential $V$ respectively correspond to disordered and ordered phase, while for $Z_\parallel < 0$, a critical 
value of the minimum of $V$ separates the disordered phase from a modulated phase with an oscillating ground state,
so that these three phases meet at the point characterized by $Z_\parallel = 0$ and by the vanishing of the minimum of $V$.

As a consequence one expects that this configuration corresponds to the tricritical Lifshitz fixed point 
and, as $Z_\parallel = 0$, the role of the term $(1/2) Z_\parallel (\partial_\parallel \phi)^2 $ is now played by the term 
$W_\parallel (\partial_\parallel^2 \phi)^2 $ 
and therefore  the scaling of the parallel and orthogonal coordinates must be different.
This leads to the introduction of two different anomalous dimensions, $\eta_{ l 2}$ and  $\eta_{l 4}$, to fully 
describe the scaling of the two point functions, $\Gamma^{(2)}(q_\bot\to 0,q_\parallel=0 ) \sim q_\bot^{2-\eta_{l 2}}$ and   
$\Gamma^{(2)}(q_\bot =0,q_\parallel \to 0 ) \sim q_\parallel^{4-\eta_{ l 4}}$. Accordingly, two different correlation lengths 
with two  critical indexes are required at criticality.

It is natural to associate the two sets of coordinates with two scales: $\kappa_\bot$, $\kappa_\parallel$,
and introduce the anomalous dimensions through  the field renormalization: $ Z_\bot \propto \kappa_\bot^{-\eta_{ l 2}}$ and 
$ W_\parallel\propto \kappa_\parallel^{-\eta_{ l 4}}$. If one connects the two scales by the  anisotropy parameter $\theta$: $\kappa_\parallel=\kappa_\bot^\theta$, 
then consistency in the scaling of the two field renormalizations  in Eq.\,(\ref{startaction}) 
requires:
\be
\theta=\frac{2- \eta_{ l 2}}{4 -\eta_{ l 4} }
\label{teta}
\ee

The scaling dimension of all the other operators are directly read from Eq.\,(\ref{startaction}) and can be expressed 
for instance in terms of $\kappa_\bot$. Then, as already seen, the dimensions of $Z_\bot$ and $W_\parallel$ are 
$[- \eta_{ l 2}]$ and $[- \theta \eta_{ l 4}]$, while the dimension
of the field $\phi$ is 
\be
D_\phi^{(m)} =\frac {  d-m +\theta (m-4+ \eta_{ l 4} )} { 2} \, ,
\label{difim}
\ee
and those of  $Z_\parallel$, $W_\bot$, $V$ are  respectively:
$[\theta(2-\eta_{ l 4})]$,\, 
$[-(2+\eta_{ l 2})]$,\,
$[d+\theta m -m]$.

One immediately notices that the above scaling dimensions are rather different from those observed for instance at the 
Wilson-Fisher fixed point which are very close to the canonical dimensions because the anomalous dimension in that case 
turns out to be very small.
The anisotropic scaling pointed out above is instead realized in proximity of the Lifshitz critical point, 
if it exists. In other words, this scaling occurs only if a fixed point solution (Lifshitz fixed point) 
of the corresponding FRG  flow equations is found. 

In this case, it is interesting to notice that, while the scaling $Z_\bot$ and $W_\parallel$ depends on the sign of  $\eta_{ l 2}$ and $\eta_{ l 4}$,
the parameter $Z_\parallel$, which vanishes at the critical point at  the mean field level, is in fact a relevant parameter  
according to its scaling dimension and, on the contrary, $W_\bot$ is irrelevant. 
Therefore one expects the full fixed point solution to be unstable with respect to small
perturbations of $Z_\parallel$ around its fixed point value. 

The particular isotropic case is easily obtained by requiring that no orthogonal coordinate is present, which means that 
the above equations must be simplified by setting $m=d$, $Z_\bot=W_\bot=0$ and $\eta_{ l 2}=0$. 
Then we are  left  with  parallel coordinates only and we can define with no ambiguity : $Z\equiv Z_\parallel$, $W\equiv W_\parallel$
and $\eta\equiv \eta_{ l 4}$. It is also convenient to reexpress the scaling dimensions in terms of the orthogonal scale $k\equiv 
\kappa_\parallel=\kappa_\bot^\theta$, instead of $\kappa_\bot$, in order to absorb $\theta$ in the scale parameter.
Then, the scaling dimensions of $W$, $Z$, $V$ become  $[- \eta]$, \,
$[2-\eta]$, $[d]$ and, from  Eq.  (\ref{difim}) , the dimension of $\phi$ is   $D_\phi =D_\phi^{(m=d)} $ :
\be
D_\phi=\frac{d-4+\eta}{2} \, .
\label{difi}
\ee
The isotropic case resembles the standard analysis where, in addition to the standard $O(\partial^2)$ kinetic term,
an additional quartic term, $O(\partial^4)$, is added to the action. However, in the standard analysis the quartic term is irrelevant
and the quadratic is marginal, while here, according to the different scaling, the quartic terms is marginal and the quadratic is relevant 
and  the occurrence  of a Lifshitz fixed point directly depends on the interplay of these two parameters.

\section{PTRG flow equation}

Once the scaling of the various operators is set, one has to look for fixed point solutions of the FRG equations and,
as already anticipated, we  make use here of the PTRG  flow equation 
\be
\label{eq:gamma}
k\;\frac{\partial \Gamma_k }{\partial k }=
-\frac{1}{2} {\rm Tr} \; \int_0^\infty \;\frac{{\rm d}s}{s} \;
k \frac{\partial g_k}{\partial k}\;\;
{\rm exp}\; \Big (-s\frac{\delta^2 \Gamma_k}{
\delta \phi\delta \phi} \Big )   
\ee
with the specific choice of the step function as a  regulator, properly adjusted for 
our quartic in the momentum two-point function:  $g_k =\Theta\left (1 - 2 W k^4 s \right )$.
A specific ansatz for the scale dependent action $\Gamma_k$ is already given in Eq.\,(\ref{startaction})
where in general it is assumed that all  parameters $W_\parallel, \,W_\bot, \, Z_\parallel, \, Z_\bot, \,V$
depend both on the field $\phi$ and on the running scale $k$, so that the renormalization effects are encoded
in the flow of these parameters with $k$. In addition, as we are interested in studying the isotropic case, 
we must set $m=d$  and  discard the parallel coordinates subspace in the action and therefore,
the full flow equation Eq.\,(\ref{eq:gamma}) in the approximation scheme of the derivative expansion, \cite{morris2},
is reduced to a  set of three coupled partial differential equations for $V$,  $W$, $Z$.

The derivation of the flow equations with terms involving four field derivatives  is rather long, 
but we can easily adapt the flow equations derived in  \cite{litimzappala}
to study  the Ising universality class at order $O(\partial^4)$ in the derivative expansion,
with the only change related to the different  scaling behavior of the various parameters at the Lifshitz critical point.
Therefore, by following \cite{litimzappala}, and  after rescaling the field and $V$,  $W$, $Z$ by their scaling dimensions : 
$\phi= k^{D_\phi}\, x$, $W(k,\phi) = k^{-\eta}\, {\it w}(k,x)$, $Z(k,\phi)= k^{2-\eta}\,{\it z}(k,x)$, $V(k,\phi) = k^{d}\, {\it v}(k,x)$,
the three flow equations read:

\be \label{vadim}
k\partial_k {\it v} - d\, {\it v}  + D_\phi ~x \partial_x {\it v}
=  \int \frac{{\rm d}^d p}{(2\pi)^d}  \; 
{\rm e}^{\left (- \frac{  a_0 }{ 2 {\it w}}  \right)} 
\ee

\be \label{wadim}
k\partial_k {\it w} + \eta   {\it w}+ D_\phi~x \partial_x {\it w}
= -\int \frac{{\rm d}^d  p}{(2\pi)^d}  \;
{\rm e}^{\left (- \frac{  a_0 }{ 2 {\it w}}  \right)} 
\;  K_{\it w}
\ee

\be \label{zadim}
k\partial_k {\it z} - (2- \eta)  {\it z} + D_\phi~x \partial_x {\it z} 
=  - \int \frac{{\rm d}^d  p}{(2\pi)^d}  
\;  {\rm e}^{\left (- \frac{  a_0 }{ 2 {\it w}}  \right)} 
\;  K_{\it z}
\ee 
where $D_\phi$ is given in Eq. \,(\ref{difi}), the parameter
$a_0=\partial_x^2  v +  z \, p^2 + 2 w\, p^4$ in the exponential stems 
from the two-point function.  $K_w$,  $K_z$ are polynomials in the loop momentum variable $p$  
respectively up to order $p^{20}$ and $p^{14}$ with coefficient functions depending on  ${\it v}, {\it w}, {\it z}$ and their first and second 
derivatives with respect to the rescaled field $x$.
The kernels $K_{\it w}$,  $K_{\it z}$  encode all the interactions among operators coming from the derivative terms of the action and
they have very long expressions which we do not report here.

Finally, a fixed point corresponds to a $k$-independent solution, $ v^* (x), {\it w^*(x)}, {\it z^*(x)}$,
of the  flow equations, (\ref{vadim},\,\ref{wadim},\,\ref{zadim}). Then, in the search for fixed points, the
first term in each of the flow equations (\ref{vadim},\,\ref{wadim},\,\ref{zadim}), involving a derivative with respect to $k$,
must be discarded and one is left with three coupled second order ordinary differential equations. 
In the scheme of the derivative expansion,
the lowest order approximation, known as Local Potential approximation (LPA),
is realized by solving Eq.  (\ref{vadim}) for $V$  and keeping fixed $w^*=1/2$ and $\eta=0$ and $z^*=0$. 
At the next order the kinetic term is turned on but,
while usually  this amounts to turning on the $O(\partial^2)$  terms and treating the  $O(\partial^4)$
as a subleading correction, in the Lifshitz  case  the leading kinetic term  involves $w^*$, the coefficient of the $O(\partial^4)$ operator.
Therefore, after discussing the LPA, we  shall first study  the coupled equations (\ref{vadim}) and (\ref{wadim}) by keeping $z=0$ 
and,  as a final step, we shall release the constraint $z=0$ and consider the full  set  (\ref{vadim},\,\ref{wadim},\,\ref{zadim}).

\section{Results}

The resolution of the set of equations (\ref{vadim},\,\ref{wadim},\,\ref{zadim}) requires a proper number of boundary conditions. In fact, 
symmetry properties of the action require vanishing of the derivatives of the solution with 
respect to the field $x$ at $x=0$: ${v^*}' (0)= {w^*}'(0)={z^*}'(0)=0$ and, in addition,
the overall normalization of the action is set by taking $w^*(0)=1/2$. Then it is important to look 
at the asymptotic behavior at large $x>>1$.

We first consider the case of positive $D_\phi>0$ in Eq. (\ref{difi}) and focus on the LPA, i.e. we fix $w^*=1/2$ and $\eta=0$ and $z^*=0$ together 
with the boundary ${v^*}' (0)=0$ and observe that the right hand side of Eq. (\ref{vadim}), in the limit of  large $x$, 
is exponentially suppressed as long as ${v^*}''(x)$ diverges in this limit. Therefore, from the left hand 
side of Eq. (\ref{vadim})  it is easy to check  that the potential diverges at large $x$ as  $v(x)\sim x^{d/D_\phi}$ as long as $D_\phi >0$.  
Incidentally, we notice that the case with a divergent potential with  power $0<{d/D_\phi}\leq 2$, such that its second derivative 
vanishes or tends to a finite value 
and the exponential does not suppress the  right hand side of Eq. (\ref{vadim}), is excluded because it would require $d<0$.

This power-law divergent behavior of the potential $v^*(x)$ at large $x$ puts a very strong constraint on Eq. (\ref{vadim}).
In fact,  only a discrete number of values $v^*(0)$ produces  a solution with  no singularity 
at any finite $x$. Any other different value of the boundary $v^*(0)$ yields a solution that is singular at some finite $x$.
By retaining only those solutions that are regular on the full $x$ axis, we find at most a discrete set of fixed point solutions.
Remarkably,  the same kind of structure is encountered and discussed in \cite{morris2,morris3} 
for the standard problem  which corresponds to setting $m=0$ in the action (\ref{startaction})
(and the anomalous dimension is $\eta_{l 2}$, while $\eta_{l 4}=0$), 
the only difference being the  scaling dimension of the field:
\be 
D^{m=0}_\phi=\frac{ (d-2+\eta_{l 2})} {2} \, .
\label{candim}
\ee
In \cite{morris3}, the only  non-singular (and non-gaussian)  solution found  for $2<d<4$ is  the Wilson-Fisher fixed point.

\begin{figure}
\includegraphics[width=10.0cm]{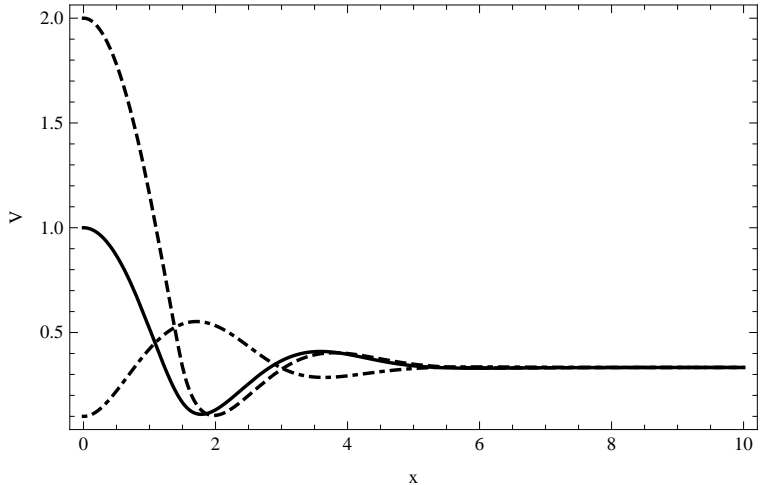}
\caption{\label{fig1}
Solutions of Eq. \, (\ref{vadim}) in the  LPA at $d=3$ and with boundaries
$v^*(0)=1$ (solid),  $v^*(0)=2$ (dashed),  $v^*(0)=0.1$ (dot-dashed), respectively. 
Note the asymptotic behavior, constant for any initial condition.
}
\end{figure}

We now include the other differential equations (\ref{wadim},\,\ref{zadim}) with the related boundaries 
${w^*}'(0)={z^*}'(0)=0$  $w^*(0)=1/2$ and again with $D_\phi>0$. 
As in the case of the LPA, where the asymptotic behavior of the solution selects a discrete number of 
values $v^*(0)$, here the same asymptotic structure  allows for a discrete number of solutions,
 i.e. of values of the 
three parameters $v^*(0)$, $z^*(0)$ and $\eta$.

Let us now examine 
the structure of the equations when  $D_\phi <0$.  
In this case the potential is
no longer divergent at large $x$ and it is expected to converge   
to a finite value. As a consequence,  in the right hand side of 
Eq. (\ref{vadim}) the exponential at large  $x$ tends to 1 and this 
determines unambiguously  the limiting value of the potential for $x\to\infty$:
$v^*(x)\to \overline v$ . Then,
even  the  subleading vanishing term in the potential  at large $x$ 
is determined from Eq.  (\ref{vadim}),  up to a constant factor $\alpha$:
$v^*(x)\sim  \overline v + \alpha \, x^{d/D_\phi}$.  
The different asymptotic behavior of the potential  drastically modifies
the spectrum of the solutions  from discrete to continuous.
In fact,  in the LPA  when $D_\phi <0$, one finds  different non-singular solutions of Eq.
(\ref{vadim})  for each value assigned to  the boundary
$v^*(0)$  (in all cases the second  boundary, ${v^*}' (0)=0$,
is to be enforced), i.e.
a line of fixed points is observed, parametrized by the  value of $v^*(0)$.
As an example, three  fixed  potentials, $v^*(x)$,  obtained  in the LPA at $d=3$
are shown in Fig. \ref{fig1}  for three different values of $v^*(0)$.

\begin{figure}
\includegraphics[width=10.0cm]{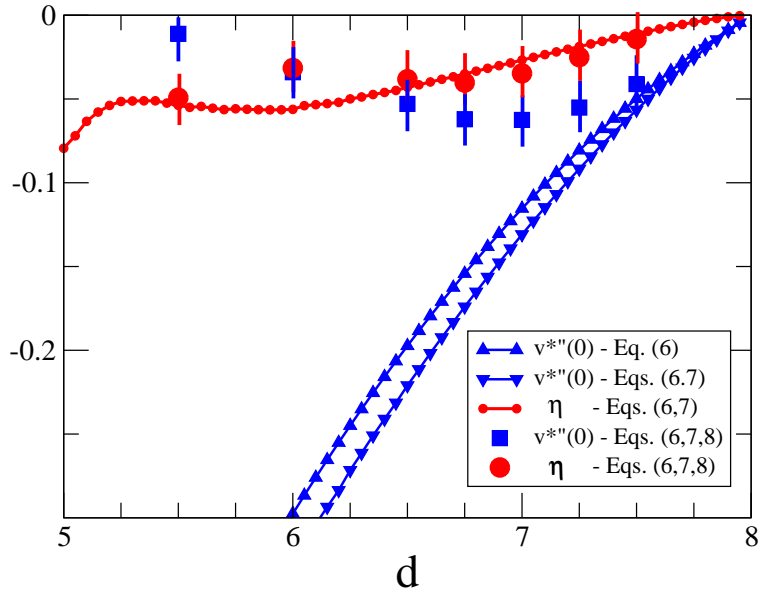}
\caption{\label{fig2} 
${v^*}''(0)$ computed in the LPA (blue triangles pointing upward), by
solving Eqs.(\ref{vadim},\ref{wadim}) (blue triangles pointing downward)
and by solving Eqs.(\ref{vadim},\ref{wadim},\ref{zadim}), (blue squares).
Anomalous dimension $\eta$ as obtained from Eqs.(\ref{vadim},\ref{wadim}) (small red circles),
and from Eqs.(\ref{vadim},\ref{wadim},\ref{zadim}) (large red circles).
(For interpretation of the references to color in this figure legend, the reader is referred to 
the web version of this article.)}
\end{figure}

If instead  $D_\phi >0$ i.e. if $d>4$, we find, by numerical
resolution of  Eq.  (\ref{vadim}),  only one non-gaussian solution 
at each fixed  $d$, and the corresponding  ${v^*}'' (0)$
is displayed  in Fig.  \ref{fig2} (blue triangles pointing upward).   
${v^*}''(0)$ grows monotonically  in the full interval $4<d<8$ although
it is  only partially visible  in Fig.  \ref{fig2}.

It is very interesting  to look at the upper and
lower critical dimensions in this problem. 
As discussed in \cite{horn},  in the  anisotropic case with $m\neq d$, 
the upper critical dimension is  
\be
\label{upper}
d_u (m)=4+ \frac{m}{2}  \, .
\ee
Eq. (\ref{upper}) in the isotropic case with $m=d$ becomes $d_u=8$, 
while when $m=0$ the known  result  $d_u=4$ is recovered.  
The value of the upper critical dimension is also obtained by requiring 
that there must be  at least one relevant interaction operator in
the potential  in order to have a non-gaussian fixed point.
In fact,  after expanding the potential in powers of the
field, the smallest interaction operator is $\lambda \phi^4$
(cubic and other odd powers are excluded because of  the 
symmetry of the action) and  $\lambda$ is relevant 
if its scaling dimension  is positive. 
Then, by recalling the dimension of the field,  Eq. (\ref{difim}),
and the definition of $\theta$,  Eq. (\ref{teta}),
it  is easy to find that  the dimension of $\lambda$ is positive if 
$d < 4 + m( 1- \theta)   -2 \eta_{ l 2} $. The upper limit, apart from 
the small corrections due to $\eta_{ l 2}$ and  $\eta_{ l 4} $, 
coincides with Eq. (\ref{upper}).
This result  confirms our numerical findings in the LPA (where the 
anomalous dimension is neglected), which 
show the presence of an isotropic Lifshitz  point only  below $d=8$.
We shall see that this result holds  even in the higher order approximation, 
where the  anomalous dimension turns out to be zero at  $d=8$.

Let us now turn to the lower critical dimension. It is known \cite{diehl},
that in the case of an $N$-component  vector field with symmetry $O(N)$,
the lower critical dimension that marks the limit below which the Goldstone 
fluctuations destroy long range order, is
\be
\label{lower}
d^{O(N)}_l=2+\frac{m}{2} \, ,
\ee
while for the Ising case,  $N=1$ and $m=0$,  the lower critical dimension becomes :
\be
\label{ising}
d^{Ising}_l=1 \, .
\ee

One immediately realizes  that 
the asymptotic behavior of the fixed point  solution of the FRG
equations  is strictly connected to the lower critical dimension.  
In fact, as discussed above for the Ising  isotropic case with $m=d$ and $N=1$,
 the nature of the solutions of Eqs. (\ref{vadim}, \ref{wadim}, \ref{zadim}),
essentially depends  on the sign of $D_\phi$, but the same argument could be repeated
 for the $O(N)$ theory with $m<d$, as the left hand side of the equations
which determines the asymptotic behavior of the solution, remains substantially unchanged.
Therefore, a discrete spectrum is obtained only if $D^{m}_\phi > 0$ which, according to
Eq. (\ref{difim}) and Eq. (\ref{teta}),  gives  
\be
d >  2+ m (1-  \theta)   -  \eta_{ l 2}  \, ,
\label{diseg}
\ee
and this  is in agreement  with Eq.  (\ref{lower}) if  $\eta_{ l 2}$ and  $\eta_{ l 4} $
are neglected.

As a check, we can restrict ourselves to the case $m=0$ where
$\eta_{ l 4}=0 $ and
we can make use of  already known  results  on  $\eta_{ l 2} $ which, as could be expected, is a function of  $d$.  
By using FRG  techniques, in \cite{codello}  it is shown  that 
$\eta_{ l 2} =0$ for $d \leq 2$ in the  $O(N)$ theory.  Therefore, 
Eq. (\ref{diseg})  reduces to  $d > 2 $ and we find full  agreement with Eq.  (\ref{lower})
at $m=0$ because of the vanishing of the anomalous dimension for $d \leq 2$.
Again with  $m=0$ but for  the Ising case, $N=1$,  
a positive  $\eta_{ l 2} > 0$  is obtained for $d \geq 2$  in \cite{codello}
which indicates that the right hand side in Eq. (\ref{diseg}) is now smaller than 2.
Unfortunately in  \cite{codello}   $\eta_{ l 2} $ is not computed 
for $d < 2$ and we can only deduce that  Eq. (\ref{diseg}) is fulfilled for $d$ strictly smaller than 2,
which is not in contradiction with Eq. (\ref{ising}), 
although full matching would require to show  that 
$(1-\eta_{ l 2}) \to 0^+ $ when $  (d-1) \to 0^+  $.

By going back to the isotropic Lifshitz case at $m=d$, we see that, with the help of 
Eq. (\ref{teta}) and by recalling the definition introduced above, $\eta=\eta_{ l 4} $,   
Eq. (\ref{diseg}) reduces to   $d >  4 -  \eta $. This means that in the LPA,
where the approximation  $\eta=0$ is used, 
a change in the spectrum of the solution  from discrete to continuous
occurs at $d =4$, as verified in  the  numerical 
analysis illustrated  above. Then, in the approximation beyond the LPA
for $N=1$,  the anomalous dimension turns out to be negative, $\eta<0$, with the 
implication that  the change in the spectrum occurs at a  larger value $d>4$, and,
accordingly, even the lower critical dimension becomes larger than 4.
 
These examples clearly show the relation between the number of dimensions $\hat d$ at which 
$v^*(x)$ changes its asymptotic behavior from divergent to finite and the lower critical dimension $d_l$.
In fact, since a physically meaningful fixed point must exist above $d_l$ and at the same time 
such solutions are absent for  $ d < \hat d$, one concludes either that $d_l= \hat d$
or, at least, $\hat d$ represents  a lower bound for $d_l$.

Finally,  we illustrate  the numerical resolution of the FRG
equations beyond LPA  in two steps. The first one consists in 
solving the two equations (\ref{vadim}, \ref{wadim}) by keeping $z^*=0$, i.e. by 
reducing the kinetic part of the action to the $O(\partial^4)$ term, with no 
$O(\partial^2)$ contribution. The second step is the resolution of the full set of equations 
(\ref{vadim}, \ref{wadim}, \ref{zadim}), including the effects of the $O(\partial^2)$ term 
which, as already noticed above, is a relevant operator and therefore its role in determining 
the fixed point solution is essential.

We found a single solution of the coupled equations (\ref{vadim}, \ref{wadim}),
i.e. a Lifshitz critical point,  when $d$ is in the range $5 \leq d< 8$ . In this approximation 
the anomalous dimension $\eta$ is  treated as a parameter to be determined 
in the resolution of  the coupled equations. Our finding for $\eta$  is plotted in 
Fig.  \ref{fig2} (small red dots) together with 
the corresponding value of ${v^*}'' (0)$ (blue triangles pointing downward).
The latter shows a small correction with respect to the LPA case,
while the former is negative  with  a non-monotonic behavior and tends to zero
when $d \to 8^-$.  Once $\eta$ is determined, we can go back to the definition of 
upper and lower critical dimension. As anticipated, $\eta=0$ at $d=8$ and therefore
$d=8$ coincides with the upper critical dimension.
On the other hand,  Eq. (\ref{diseg})  now gives   $d >  4 -  \eta   \gta  4 $,  because  $\eta <0$,
and  $|\eta| \approx O(10^{-1})$.

\begin{figure}
\includegraphics[width=10.0cm]{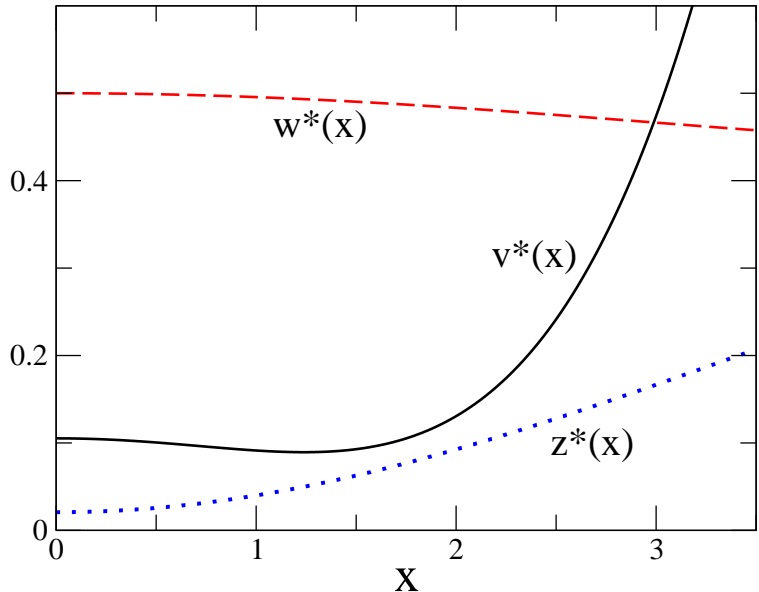}
\caption{\label{fig3} 
Solution of the coupled flow equations (\ref{vadim}, \ref{wadim}, \ref{zadim}),
$v^*(x)$ (black solid), $w^*((x)$ (red dashed) and $z^*(x)$ (blue dotted),
computed at $d=7.5$.
(For interpretation of the references to color in this figure legend, the reader is referred to 
the web version of this article.)}
\end{figure}

The final step involves the resolution of the three coupled equations, 
(\ref{vadim}, \ref{wadim}, \ref{zadim}),  and in this case the numerical 
analysis is much more demanding  and the accuracy of the results  is 
reduced by a residual dependence on the endpoint of the integration range of the field $x$. 
Therefore we solved the equations only for a few values of $d$.
A plot of the solution obtained at $d=7.5$ is reported in Fig. \ref{fig3}
while the anomalous dimension $\eta$ (large red  circles) 
and the second derivative of the potential  ${v^*}'' (0)$ (blue squares)
are reported in Fig.  \ref{fig2}  together with the estimated 
error on these quantities.

The plots in Fig.  \ref{fig2} show the importance of including all relevant parameters, such as $z^*$
in the determination of the  Lifshitz fixed point. In fact,  while turning on $z$ does not change 
the order of  magnitude of the anomalous dimension,  one observes a drastic change in  ${v^*}'' (0)$
if compared to the previous approximations. In addition, no solution was found at $d=5$, 
which suggests that the Lifshitz fixed point is effectively destroyed by the fluctuations induced by $z$
when $d$ is decreased, although  we cannot exclude a possible failure in the numerical search of the 
solution.

\section{Conclusions}

In conclusion, we have studied the isotropic Lifshitz critical behavior
 for a single component field theory, i.e. with $m=d$ and $N=1$,
by means of the PTRG flow equations. In particular we solved the fixed 
point equations first in the lowest order approximation, the LPA, and then 
in the first and second order of the derivative expansion, by including  fluctuations 
associated to the $O(\partial^4)$ and to the $O(\partial^2)$ operators.

From the constraints on the  asymptotic structure of the solution, 
already in the LPA it is evident that a single physically meaningful fixed point solution 
can be obtained only for $d > 4$ which can be related to  the lower critical dimension and, when 
the constraint  coming from of the upper critical dimension is also included,
one gets $4<d<8$.  This, on one hand, supports the Monte Carlo analysis performed  
in \cite{schmid} at $d=3$ but, on the other hand, strongly questions the reliability of the  results 
on the Lifshitz critical behavior  observed  in \cite{schwahn} at $d=3$.  

The numerical analysis performed by including the parameter $z^*$ , which is a relevant operator
that strongly influences the structure of the solution, shows the existence of a Lifshitz point within 
the interval  $5.5 < d < 8$, and the anomalous dimension $\eta$ determined at this critical points 
is always negative and $|\eta|<<1$.
In particular, no evidence of a solution in $d\leq 5$ has been found.

A final comment concerns the importance of extending this analysis to the $O(N)$ theory.
In fact, if the Lifshitz critical point survives down to $d=4$  (with $\eta=0$ at $d=4$), then, 
the lower critical dimension $d^{O(N)}_l=4$ would play for the Lifshitz case the same role 
of $d=2$ for the standard critical behavior of  the $O(N)$ theory.
\\[2ex]

\vfill\eject

\noindent{\bf Acknowledgements}\step

The authors thank   A. Codello  and H. Diehl
for e-mail correspondence.

\bibliographystyle{apsrev}
\bibliography{isotropic}

\end{document}